\documentclass{aa}
\usepackage{epstopdf}
\usepackage{txfonts}
\bibpunct{(}{)}{;}{a}{}{,} 

\begin{document}

\title{A Dusty M5 Binary in the $\beta$ Pictoris Moving Group}

\author{David R. Rodriguez\inst{1}
\and B.\ Zuckerman\inst{2}
\and Jacqueline K.\ Faherty\inst{3,4}
\and Laura Vican\inst{2}}

\institute{
Departamento de Astronom\'ia, Universidad de Chile, Casilla 36-D, Santiago, Chile 
\email{drodrigu@das.uchile.cl} 
\and
Dept. of Physics \& Astronomy, University of California, Los Angeles 90095, USA
\and
Dept. of Terrestrial Magnetism, Carnegie Institution of Washington, 5241 Broad Branch Road NW, Washington, DC 20015, USA
\and
Hubble Fellow
}

\abstract
{
We report the {identification} of a new wide separation binary (LDS~5606) in the $\sim$20~Myr-old $\beta$ Pic moving group. This M5+M5 pair has a projected separation of 26$''$, or $\sim$1700~AU at a distance of 65~pc. Both stars host warm circumstellar disks and many strong hydrogen and helium emission lines.
Spectroscopic observations reveal signatures of youth for both stars and on-going mass accretion in the primary.
The properties of LDS~5606 make it an older analog to the $\sim$8~Myr TWA~30 system, which is also composed of a pair of widely separated mid-M dwarfs, each hosting their own warm circumstellar disks.
LDS~5606 joins a rather exclusive club of only 3 other known stellar systems where both members of a binary, far from any molecular cloud, are orbited by detected {circumstellar disks}.
}



\maketitle

\section{Introduction}

Over the last few decades, astronomers have identified hundreds of stars in young moving groups near Earth (for reviews see \citealt{ZS04,Torres:2008,Malo:2013}). These moving groups have ages of $\sim$10--100~Myr and provide a sample of stars useful for the study of the planet formation environment. Previous work has relied heavily on identification of these stars at X-ray wavelengths with the ROSAT All-Sky Survey (RASS).
Recent work has demonstrated that low-mass members can be successfully identified by utilizing the GALEX \citep{Martin:2005} ultraviolet survey \citep{Rodriguez:2011,Rodriguez:2013,Shkolnik:2011}. 

Among known low-mass youthful stars within 100 pc of Earth, a modest number stand out with special interest because of the presence of both surrounding dust and gas. The first such star to be recognized was the late-K type star TW Hya; the first known member of the $\sim$8~Myr-old TW Hya Association (TWA; see \citealt{Ducourant:2014} and references therein). 
Relatively recently, \citet{Looper:2010a,Looper:2010b} identified a remarkable pair of dusty mid-M type stars (TWA 30A \& B) that are members of TWA and that exhibit surrounding gas that is both accreting and outflowing. 
As part of the GALEX Nearby Young-Star Survey (GALNYSS; \citealt{Rodriguez:2013}) we identified a pair of mid-M dwarfs that exhibit strong UV emission and excess mid-infrared emission. As we show in Sect.~\ref{results}, this very dusty binary, {LDS~5606}, is a likely member of the $\sim$20~Myr-old $\beta$~Pic moving group and is an older analog of the TWA~30 binary.
Dusty low-mass stars such as these serve as excellent laboratories in which to study the planet forming environment of the most abundant stars in the Galaxy.

To date, only a handful of binaries far from molecular clouds or star forming regions are known in which both stars host circumstellar disks. 
LP~876--10, recently recognized as a widely separated companion (158~kAU) to Fomalhaut \citep{Mamajek:2013}, has now been shown to host a cold circumstellar disk \citep{Kennedy:2014}. The HD223352 triple system hosts a circumbinary disk around the close binary and a circumstellar disk around the $\sim$3000~AU companion, HD223340 \citep{Phillips:2011}. Both systems are older than 100~Myr.
At the intermediate age range ($\sim$10--100~Myr) of most nearby young moving groups, only TWA~30 is a known binary hosting a pair of disks. The TWA~30 system is comprised of an M4+M5 pair separated by $\sim$3400~AU \citep{Looper:2010a,Looper:2010b}.
Among star forming regions, there are many more cases where both stars in a binary system each host disks. \citet{Monin:2007} 
conservatively list 40 multiples with pairs of disks (either accreting or not), but argue that more non-accreting disks pairs remain to be discovered in these $\sim$few~Myr-old regions.
It is clear that, while binary systems do host disks, these dissipate on faster timescales than around single stars \citep[e.g.,][]{Rodriguez:2012}. The presence of two disks within a $\gtrsim$8~Myr binary is a rare occurrence; LDS~5606 presented here is only the 4th example after Fomalhaut, HD~223352, and TWA30.

\begin{figure}[t]
\begin{center}
\includegraphics[width=88mm,angle=0]{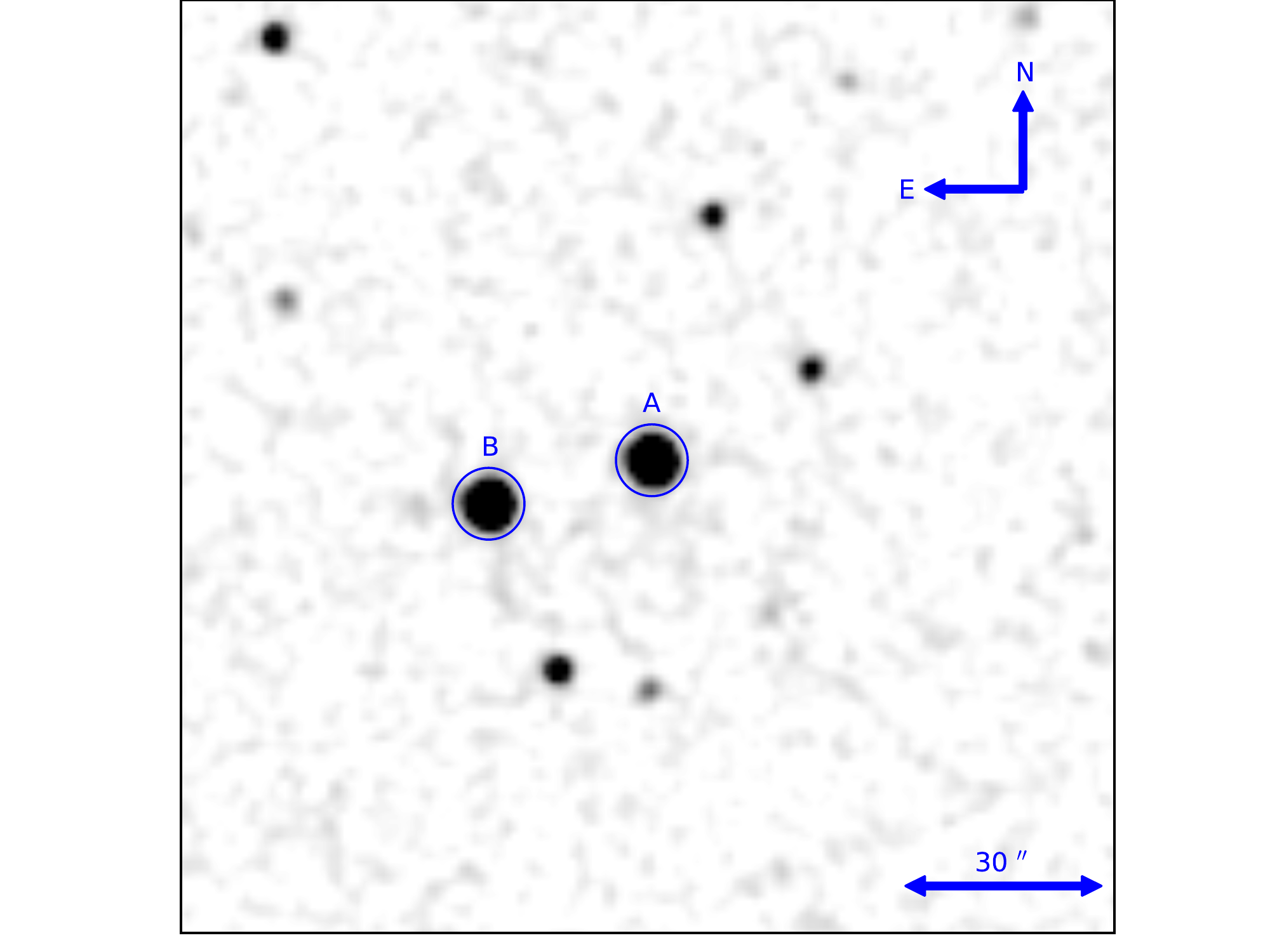}
\end{center}
\caption{2MASS J-band finder chart illustrating the location {and components of LDS~5606}.}
\label{fig:chart}
\end{figure}

{LDS~5606 is a 26\arcsec\ binary listed in the Luyten Double Star catalog \citep{Luyten:1997}. Other than an entry in the Washington Double Star catalog \citep{Mason:2001}, this system has gathered little attention since its discovery. 
Given that both stars have nearly identical 2MASS magnitudes, spectral types, and similar UV and IR excesses, it is difficult to unambiguously designate one or the other as the brighter primary.
Hence, we adopt the notation introduced in the LDS and WDS catalogs that the primary star (A) is the western component of the binary (see Fig.~\ref{fig:chart}).
Both stars are detected in GALEX yet, curiously, the secondary component (B) was not identified as part of GALNYSS given that its near UV (NUV) emission is stronger than the selection criteria (see \citealt{Rodriguez:2013}), a situation similar that of the actively-accreting TW~Hya.}
In this paper, we demonstrate that LDS~5606 is a new member of the $\beta$~Pic moving group. In a related paper (Zuckerman et al., submitted to ApJ), we discuss in more detail the disk and accretion signatures of both stars.

\section{Observations}

\subsection{Magellan-MagE}

On 2013 January 31 (UT) we used the 6.5m Clay Magellan telescope and Magellan Echellette Spectrograph (MagE; \citealt{Marshall:2008}) to obtain optical spectra of LDS~5606A.  MagE is a cross--dispersed optical spectrograph, covering 3,000 to 10,000~\AA~ at medium resolution.  Our observations employed a $0.7^{\prime\prime}$ slit aligned at the parallactic angle (resolution $\lambda$/$\Delta \lambda \sim$ 4000). Observations were made under clear conditions with an average seeing of $\sim$0.6$\arcsec$.  At the time, we had not identified LDS~5606B so no spectrum was acquired on that source.  A 300s integration was obtained for LDS~5606A followed immediately by a 3s ThAr lamp spectrum for wavelength calibration.  The spectrophotometric standard GD~108 was observed for flux calibration (120s). Ten Xe-flash and Quartz lamp flats as well as twilight flats were taken at the start of the evening for pixel response calibration. The data were reduced using the MagE Spectral Extractor pipeline (MASE; \citealt{Bochanski:2009}) which incorporates flat fielding, sky subtraction, and flux calibration IDL routines.

\subsection{IRTF-SpeX}
We obtained low resolution near-infrared spectroscopy of both components using the SpeX spectrograph (\citealt{Rayner:2003}) mounted on the 3m NASA Infrared Telescope Facility (IRTF).   On 2013 February 28 (UT), we used the spectrograph in prism mode with the 0.8$\arcsec$ slit aligned to the parallactic angle. This resulted in $\lambda$ / $\Delta\lambda~\approx$~100 spectral data over the wavelength range of 0.8--2.4 $\mu$m. Conditions included light cirrus and the seeing was 0.8$\arcsec$ at $K$. We observed LDS~5606A followed by LDS~5606B.  Our strategy employed 4 individual exposures of 90 seconds in an ABBA dither pattern along the slit.

\begin{figure}[t]
\begin{center}
\includegraphics[width=88mm,angle=0]{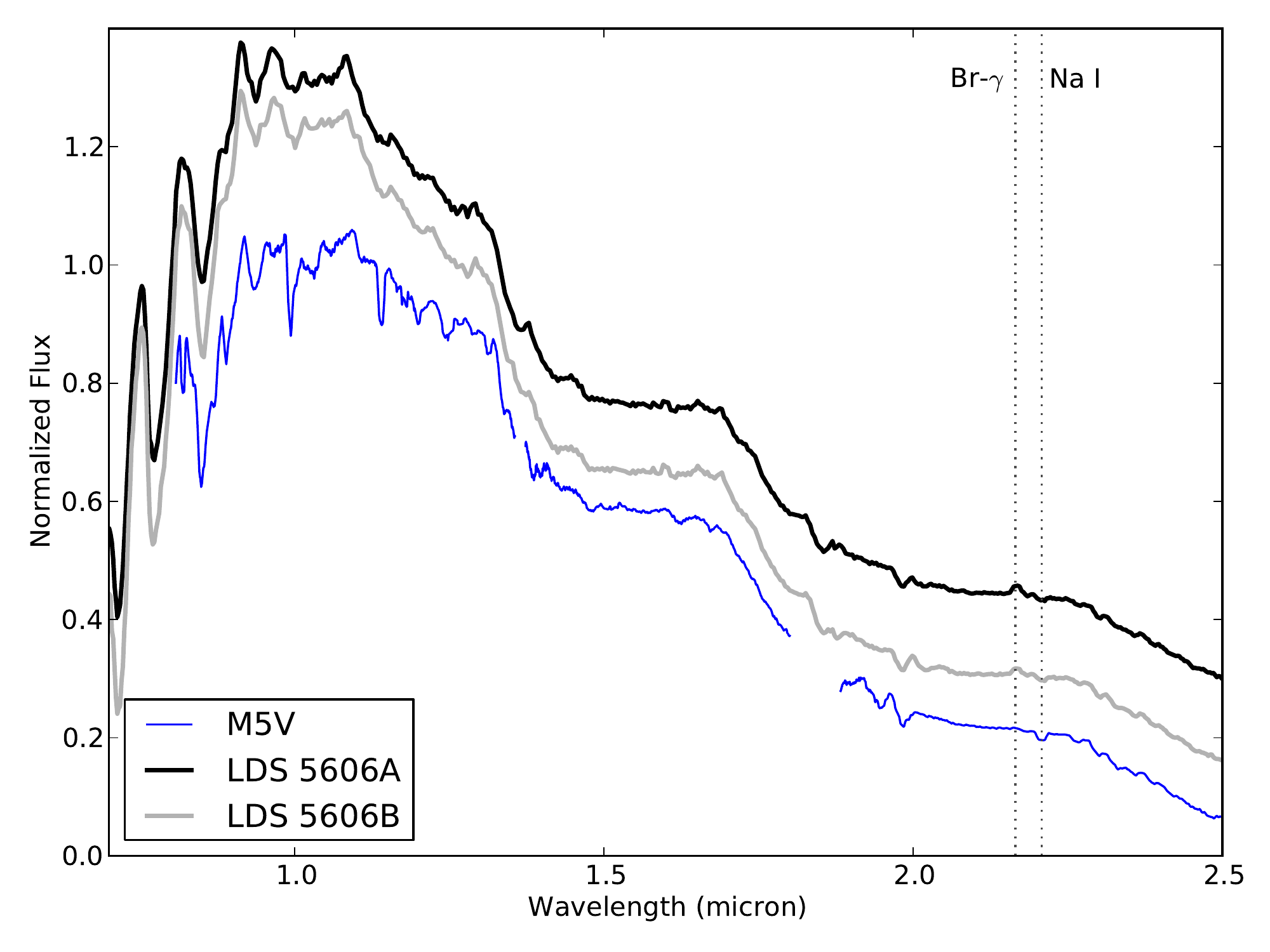}
\end{center}
\caption{IRTF-SpeX prism spectra of LDS~5606 compared to an M5 standard. Both stars show weak Na~I absorption at 2.2$\mu$m and some possible Br-$\gamma$ emission.}
\label{fig:spex}
\end{figure}

Immediately after the science observation we observed the A0V star HD35036 at a similar airmass for telluric corrections and flux calibration. Internal flat-field and Ar arc lamp exposures were acquired for pixel response and wavelength calibration, respectively. All data were reduced using the SpeXtool package version 3.4 using standard settings (\citealt{Cushing:2004}, \citealt{Vacca:2003}). Figure~\ref{fig:spex} shows the spectra compared to an M5 standard.

\subsection{Keck-HIRES}

We used the HIRES echelle spectrometer \citep{Vogt:1994} on the Keck I telescope at Mauna Kea Observatory in Hawaii. 
Spectra were obtained during observing runs in 2013 October and November. 
The red cross disperser was combined with a 1.15$''$ slit aligned at the parallactic angle and the wavelength range between 4370 and 9000\AA\ was covered with a resolution of $\sim$40,000. 
Spectra were reduced using the IRAF and MAKEE software packages. 

LDS~5606A was observed on October 17 and November 16. The echelle grating settings were not the same in October and November so the wavelength coverage differed at the extreme wavelengths.
Clouds were present during both nights and hampered the integrations (which were terminated early). Integration times were 1900 and 1713 sec on October 17 and November 16, respectively.
LDS~5606B was observed on 2013 October 21 for 2300 sec. Some thin clouds were present on this night. 

{By measurement of centroids of absorption lines we determine} radial velocities of 
$14.2\pm0.9$~km/s and $14.6\pm0.3$ for LDS~5606A and B, respectively. 
To establish the RV scale, two of the three RV standards HIP~26689, 47690, and 117473 \citep{Nidever:2002} were observed on each of the three nights of HIRES observations.
We also cross-correlated the spectra against the radial velocity standard HIP~117473 of M2 spectral type with IRAF's {\it fxcor} task to measure 
$15.3\pm0.7$ and $15.6\pm0.7$~km/s for LDS~5606A and B, respectively.

\subsection{VLT-UVES}

\begin{figure*}[t]
\begin{center}
\includegraphics[width=17cm,angle=0]{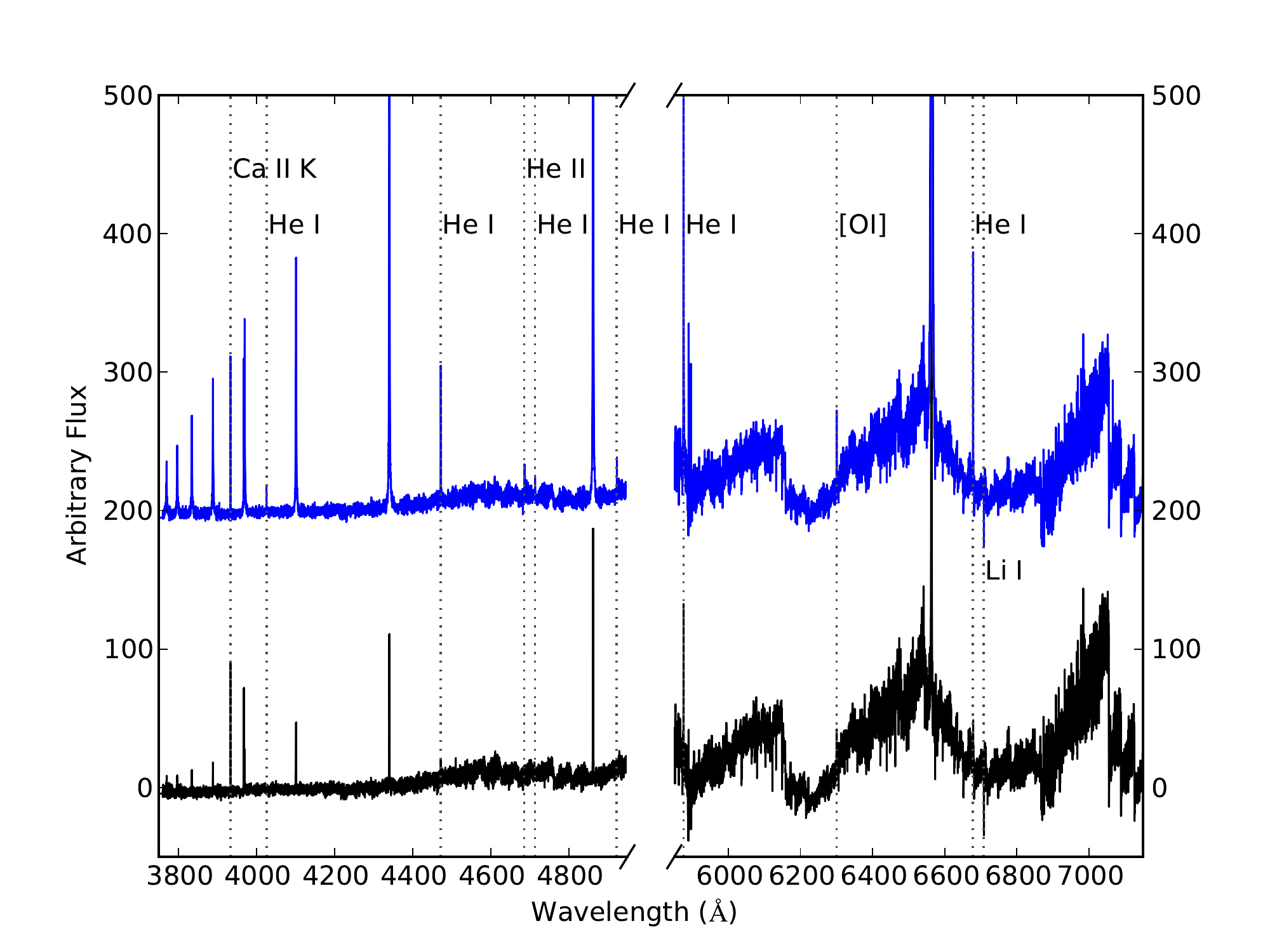}
\end{center}
\caption{VLT-UVES spectra for LDS~5606A (top) and LDS~5606B (bottom), smoothed and arbitrarily scaled for clarity. Many prominent emission lines are evident, particularly for LDS~5606A. Emission from the core of the Na-D double at 5890 and 5896\AA, and Ca~II H at 3968\AA\ (blended with H$\epsilon$ at 3970\AA) is not labeled. All other strong, unlabeled emission lines are from the Hydrogen Balmer series, which can be seen to extend from n=3 (H$\alpha$) to n=11. Expanded portions of this spectrum highlighting a variety of spectral features, including comparison with portions of the HIRES spectra, appear in Zuckerman et al.\ (submitted to ApJ).}
\label{fig:spec}
\end{figure*}

Both components of LDS~5606 were observed on 2013 December 3 (UT) with the UVES spectrograph \citep{Dekker:2000} on the Very Large Telescope (VLT).
Observations were carried out with the standard DICHROIC2 setup with CD\#2 and CD\#4 set at 4370 and 7600\AA. This provides coverage of 3730 to 5000 and 5650 to 9460\AA\ at a resolution of R$\sim$20,000 with a 2.1$''$ slit aligned at the parallactic angle (Fig.~\ref{fig:spec}). The integration time was 900 seconds.
Data were reduced with the ESO Reflex pipeline.

To determine radial velocities, standards (HIP103039 and HIP5643) from \citet{Nidever:2002} were observed. 
We cross-correlated our observations of LDS~5606 against the standards using IRAF's {\it fxcor} task to measure radial velocities of $15.1\pm1.2$~km/s for LDS~5606A and $15.9\pm0.7$~km/s for LDS~5606B.
Although the standards were taken on separate nights (2013 Oct 9 \& 23), comparison against each other reveal close agreement (within $\sim$0.1~km/s) with the literature value.

\section{Results} \label{results}

\begin{table*}
\begin{center}
\begin{tabular}{lccl}
\hline
Parameter & LDS~5606A & LDS~5606B & Ref. \\
\hline
\hline
WISE Designation & J044800.86+143957.7 & J044802.59+143951.1 & 1\\
$\mu_\alpha$ (mas/yr)$^a$ & $24.6\pm5$ & $24.3\pm5$ & 2,3\\
$\mu_\delta$ (mas/yr)$^a$  & $-42.3\pm4$ & $-41.5\pm4$ & 2,3\\
RV (km/s)$^b$  	& $14.9\pm0.8$ & $14.9\pm0.4$ & 4\\
Distance (pc)$^c$ 	& $65\pm6$ & $65\pm6$ & 4\\
Spectral Type & M5$\pm$1 & M5$\pm$1 & 4\\
U (km/s) 	& $-12.4\pm0.9$ 	& $-12.4\pm0.6$ & 4\\
V (km/s) 	& $-16.0\pm1.9$ 	& $-15.7\pm1.9$ & 4\\
W (km/s) 	& $-6.3\pm1.4$ 	& $-6.3\pm1.4$ & 4\\
X (pc) 	& \multicolumn{2}{c}{$-61.2$} & 4\\
Y (pc) 	& \multicolumn{2}{c}{$-4.7$} & 4\\
Z (pc) 	& \multicolumn{2}{c}{$-21.2$} & 4\\
GALEX FUV (mag) & $19.98\pm0.17$ & $19.97\pm0.17$ & 5\\
GALEX NUV (mag) & $19.97\pm0.13$ & $19.26\pm0.08$ & 5\\
V (mag)$^d$ & 16.7 & 16.6 & 2,6\\
2MASS J (mag) & $11.68\pm0.02$ & $11.68\pm0.02$ & 7\\
2MASS H (mag) & $11.06\pm0.02$ & $11.07\pm0.02$ & 7\\
2MASS K (mag) & $10.73\pm0.02$ & $10.68\pm0.02$ & 7\\
WISE 3.4$\mu$m (W1 mag) & $10.22\pm0.02$ & $10.52\pm0.02$ & 1\\
WISE 4.6$\mu$m (W2 mag) & $9.77\pm0.02$ & $10.16\pm0.02$ & 1\\
WISE 12$\mu$m (W3 mag) & $7.98\pm0.02$ & $8.11\pm0.02$ & 1\\
WISE 22$\mu$m (W4 mag) & $6.18\pm0.06$ & $6.09\pm0.05$ & 1\\
$m_{bol}$ (mag) & $13.64\pm0.02$ & $13.62\pm0.02$ & 4 \\
log $L_{bol}/L_\odot$ & $-1.93\pm0.09$ & $-1.93\pm0.09$ & 4 \\
\hline\end{tabular}
\caption{Notes: ($a$) average of proper motions given in PPMXL and UCAC4.
($b$) The listed radial velocity is the weighted average of all measurements.
($c$) The listed distance is from the BANYAN kinematic analysis \citep{Malo:2013}. At 65 pc, the plane of the sky separation of the two stars is $\sim$1700~AU.
($d$) We assume V-band magnitude uncertainties of $\sim$0.3~mag.
References: (1) WISE \citep{Wright:2010}; (2) UCAC4 \citep{Zacharias:2012}; (3) PPMXL \citep{Roeser:2010}; (4) this work; (5) GALEX GR6/7 \citep{Martin:2005}; (6) NOMAD \citep{Zacharias:2004}; (7) 2MASS \citep{Cutri:2003}. } \label{tab:summary}
\end{center}
\end{table*}

\begin{figure}[t]
\begin{center}
\includegraphics[width=88mm,angle=0]{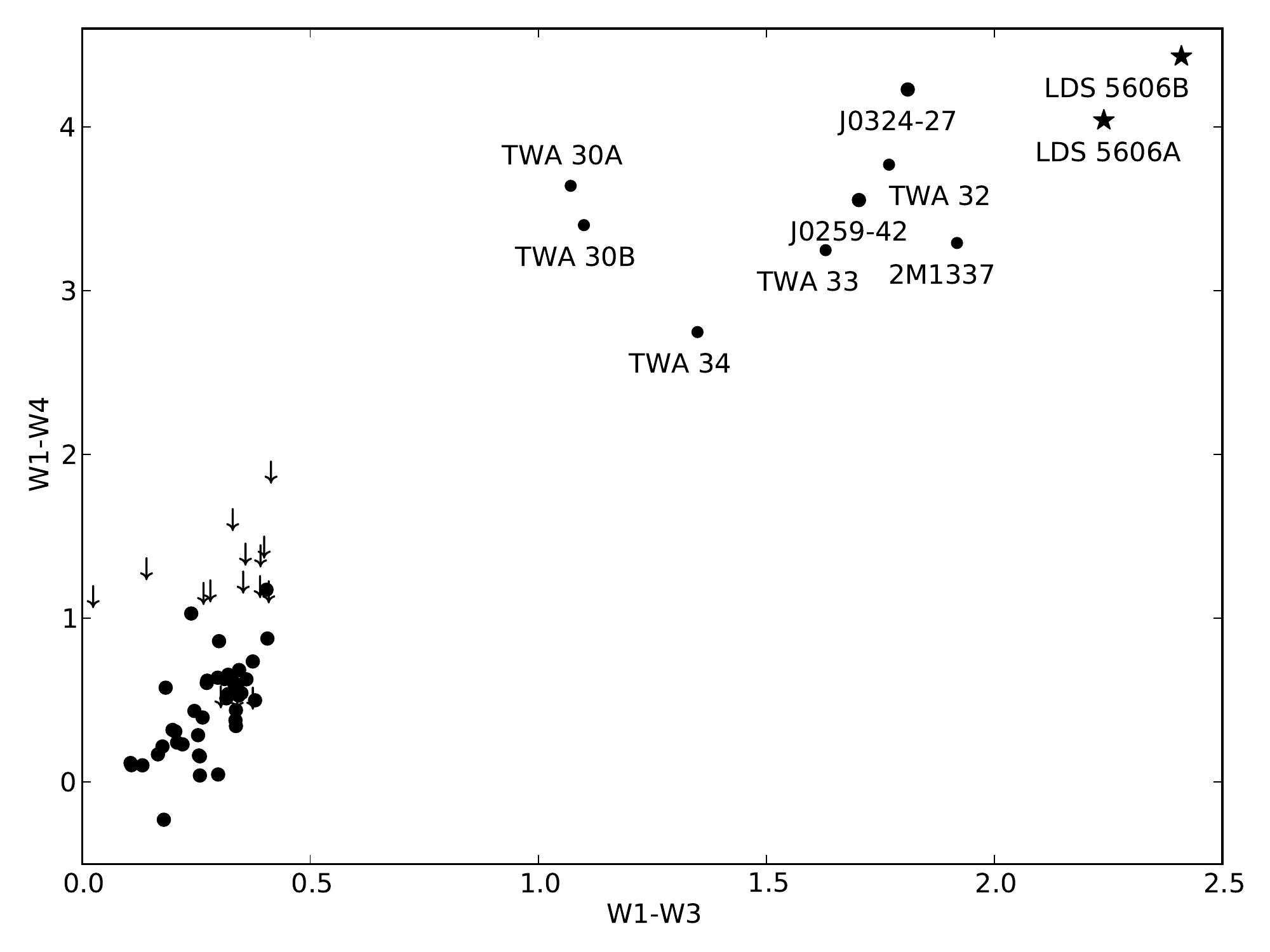}
\end{center}
\caption{W1--W3 and W1--W4 colors for early to mid M-dwarfs from \citet{Rodriguez:2013} as well as labeled dusty M-dwarfs  \citep{Looper:2010a,Looper:2010b,Schneider:2012a,Schneider:2012b}. The colors of TWA~30B have been modified as described in \citet{Schneider:2012a}. Both components of LDS~5606 exhibit very red colors comparable to those of stars known to host warm circumstellar dust disks. }
\label{fig:irexcess}
\end{figure}

LDS~5606 was identified as part of GALNYSS (GALEX Nearby Young-Star Survey; \citealt{Rodriguez:2013}) and immediately flagged as an interesting system; Table~\ref{tab:summary} summarizes its parameters.
The secondary component had such blue NUV-W1 colors that it was not picked up as part of the survey, but was recovered because of its common motion to LDS~5606A (see finding chart in Fig.~\ref{fig:chart}). Both stars show evidence of WISE IR excess by their red W1--W3 and W1--W4 colors (Fig.~\ref{fig:irexcess}). Their proper motions, as we describe below, place them as candidate members for the $\beta$~Pic moving group.

Utilizing the convergent point tool in \citet{Rodriguez:2013} and BANYAN tool in \citet{Malo:2013}, we find $\beta$~Pic membership probabilities of 66 and 79\%, respectively. These kinematic tools also predict radial velocities of $\sim$14.5~km/s and distances of $\sim$65~pc.
At a distance of 65~pc, both components would have $M_K\sim$6.6. With V-band magnitudes from UCAC4 and NOMAD, we find V--K is $\sim$6, consistent with an M5 spectral type. An M5 with this $M_K$ matches well the empirical $\beta$~Pic isochrone in \citet{ZS04}. 
For a spectral type of M5, an old field dwarf would have $M_J\sim9.6$ ($M_K\sim8.6$; \citealt{Cruz:2002}), which, at a distance of 65~pc, would suggest an apparent magnitude of J$\sim$13.7 (K$\sim$12.7). LDS~5606 is more luminous than these estimates, consistent with a young age (see Sect.~\ref{youth}).
In consideration of group membership, we implicitly assumed that both members of the visual binary are single stars. Should both stars actually be as yet unresolved (and thus unrecognized) roughly equal luminosity binaries, then $M_K$ would be inconsistent with $\beta$~Pic membership. Our high-resolution UVES and HIRES spectra show no evidence of binarity.

\begin{figure}[htb]
\begin{center}
\includegraphics[width=88mm,angle=0]{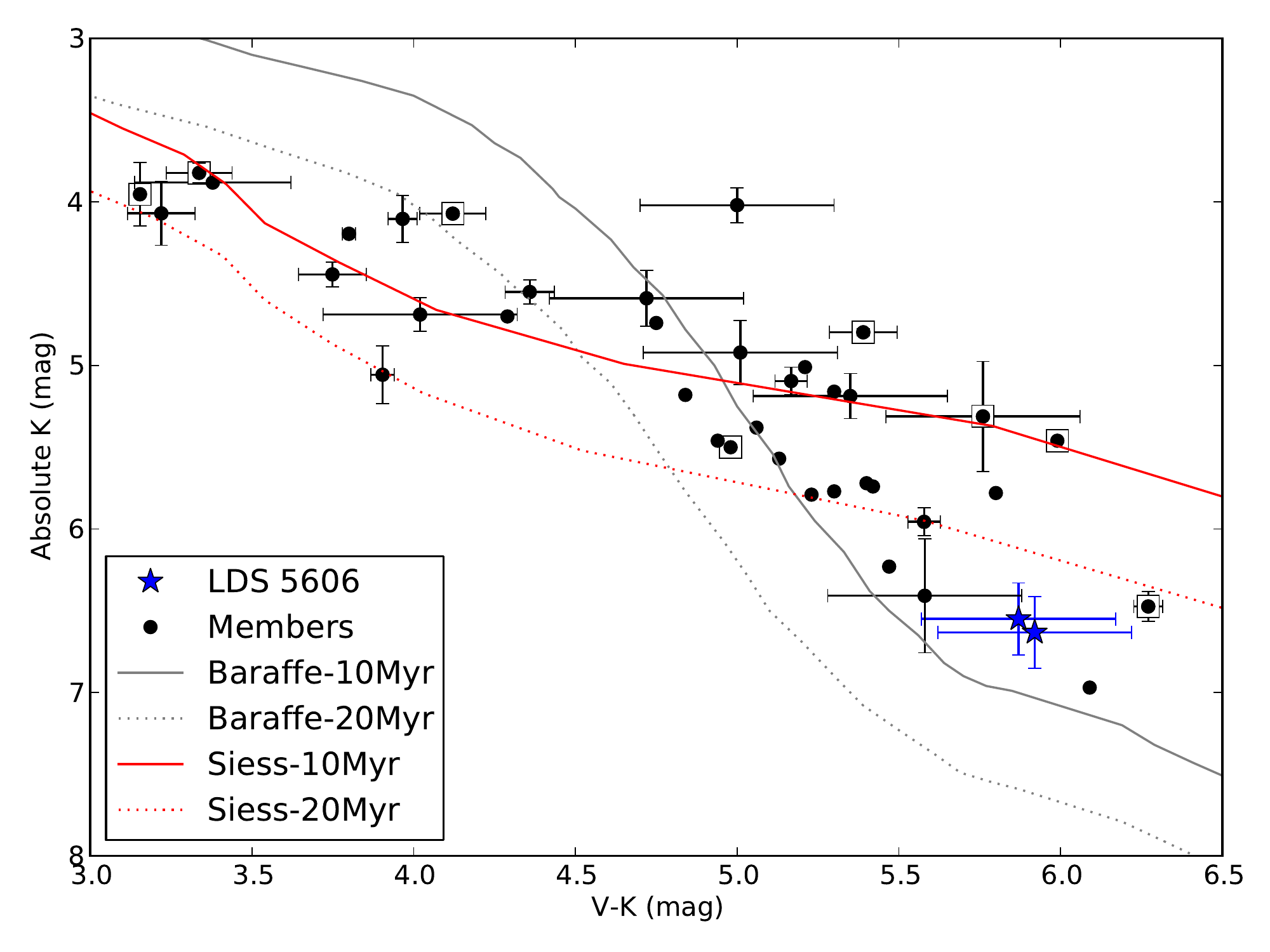}
\end{center}
\caption{Color-magnitude diagram of $\beta$~Pic moving group members (as drawn from \citealt{Malo:2013,Binks:2013,Riedel:2014}) and the LDS~5606 binary system. Unresolved binaries are indicated with squares. For the sample from \citet{Riedel:2014}, we used their deblended magnitudes without uncertainties (see their Table~5). V-band magnitudes for LDS~5606 are drawn from UCAC4 and NOMAD; we assume an uncertainty of 0.3 magnitudes for these faint stars. Theoretical isochrones from \citet{Baraffe:1998} and \citet{Siess:2000} are shown for comparison.}
\label{cmd}
\end{figure}

Figure~\ref{cmd} shows a color-magnitude diagram of $\beta$~Pic moving group members (from \citealt{Malo:2013,Binks:2013,Riedel:2014}) and the LDS~5606 system along with theoretical 10 and 20~Myr isochrones \citep{Baraffe:1998,Siess:2000}. 
The two models differ substantially at these late spectral types, which makes identification of additional late-type members of moving groups invaluable to better constrain stellar evolution models.

With our UVES spectra, we measured the TiO5 index \citep{Reid:1995} to estimate the spectral type of both stars, accurate to 0.5 spectral types. 
We measure M4.7 and M5.4 for LDS~5606A and B, respectively.
For the HIRES data, a gap between orders falls at the TiO5 {wavelengths rendering this index unusable.}
A by-eye comparison of the MagE, HIRES, and UVES spectra with active M-dwarf templates from \citet{Bochanski:2007} suggest a spectral type of $\sim$M5 is appropriate.
In the near-IR, both stars appear to be M5 either by eye or by utilizing some of the spectral type indices summarized in Table~3 of \citet{Allers:2013}.
We thus adopt spectral types of M5$\pm$1 for both components of LDS~5606.
\citet{Pecaut:2013} suggest a $T_{eff}$ of 2880~K for young (5--30~Myr) M5 dwarfs. 
{With the bolometric corrections listed in \citet{Pecaut:2013} and the 2MASS magnitudes, we estimate a bolometric luminosity of 0.012~$L_\odot$ for each component. The \citet{Baraffe:1998} models suggest a $\sim$20~Myr-old object with that luminosity would have a mass of 0.13~$M_\odot$ (0.09~$M_\odot$ if $\sim$10~Myr-old). 
The \citet{Siess:2000} models suggest similar masses. However, the temperatures these models predict is warmer (3170~K) than suggested for young M5 dwarfs. Matching a cooler temperature, however, suggests a mass of $\sim$0.06~$M_\odot$ for each component of LDS~5606 regardless of whether we use the 10 or 20-Myr model. This temperature (and color) discrepancy has been noted before (e.g., see \citealt{Pecaut:2013}). Hence, we adopt a mass of $\sim$0.1~$M_\odot$ for both components.
}

At a statistically de-projected separation of $1.26\times1700\approx2140$~AU \citep{Fischer:1992} and mass of 0.1~$M_\odot$, the binding energy of this wide binary is $E=G M_1 M_2 / a = 0.8 \times 10^{41}$~ergs. 
This binding energy is comparable to some of the other $\gtrsim$8~Myr disk binary systems, notably TWA~30AB ($0.9\times10^{41}$; \citealt{Looper:2010b}) and Fomalhaut A--C ($0.4\times10^{41}$; \citealt{Mamajek:2013}). The HD~22352/40 A0+K1 system is more tightly bound ($115\times10^{41}$; \citealt{Phillips:2011}).
Additional binaries with comparable (low) binding energies can be seen in Figure~8 of \citet{Faherty:2011}.

{Assuming a circular orbit, the two components of LDS~5606 would orbit each other with a period of $\sim$221,000 years and orbital velocity $\sim$0.3~km/s. At a distance of 65~pc, this amounts to a motion of $\sim$0.9~mas/yr. While small, this motion can be detected by the GAIA satellite. With the V--Ic color typical of young M5 dwarfs (3.31; \citealt{Pecaut:2013}) and the G--V relationship in \citet{Jordi:2010}, we estimate a G magnitude of $\sim$15 for LDS~5606. For a G$\sim$15 star, GAIA is expected to reach a parallax error of $\sim$27$\mu$as and a proper motion error of $\sim$14$\mu$as/yr \citep{deBruijne:2012}\footnote{See GAIA's Science Performance at \url{http://www.cosmos.esa.int/web/gaia/science-performance}}. 
However, while 900~$\mu$as/yr motion should be easily detectable with GAIA, the baseline will likely be insufficient to characterize the orbit or estimate dynamical masses.}

\subsection{Youth Features} \label{youth}

\begin{table*}
\begin{center}
\begin{tabular}{llcrrr}
\hline
Object & Instr.\  & UT Date & H$\alpha$ EW (\AA) & Li EW (\AA) & Na I Index \\
\hline
\hline
LDS~5606A	& MagE & 2013-1-3 & $-73.6\pm3.1$ & $0.389\pm0.055$ & 1.16 \\
	& HIRES & 2013-10-17 	& $-99.4\pm7.9$ & $0.382\pm0.041$ &  \\
	& HIRES & 2013-11-16 	& $-135.3\pm10.2$ & $0.380\pm0.042$ & \\
	& UVES & 2013-12-3	&	$-84.9\pm7.8$	&	$0.383\pm0.054$	&	1.16	\\
\hline
LDS~5606B	& HIRES  & 2013-10-21 & $-25.1\pm2.8$ & $0.501\pm0.047$ &  \\
	& UVES & 2013-12-3	&	$-16.0\pm2.0$	&	$0.489\pm0.053$	&	1.18	\\
\hline\end{tabular}
\caption{Spectroscopic measurements for LDS~5606. 
The Na~I index is the ratio of the average flux on and off the doublet (see \citealt{Lawson:2009}). } \label{tab:meas}
\end{center}
\end{table*}

\begin{table*}
\begin{center}
\begin{tabular}{lcccccc}
\hline
 & \multicolumn{3}{c}{Spectral Type Indices} & \multicolumn{3}{c}{Gravity Indices} \\
Object & $H_2O$ & $H_2O$--1 & $H_2O$--2 & FeH$_z$ & K$I_J$ & $H$-cont \\
\hline
\hline
LDS~5606A	& 5.1 & 4.2 & 5.0 & $1.020\pm0.007$ & $1.006\pm0.004$ & $1.011\pm0.006$ \\
LDS~5606B	& 5.3 & 4.4 & 5.0 & $1.031\pm0.006$ & $1.008\pm0.003$ & $1.011\pm0.006$ \\
\hline\end{tabular}
\caption{IRTF-SpeX measurements for LDS~5606 of the spectral type and gravity indices listed in \citet{Allers:2013}. } \label{tab:spex}
\end{center}
\end{table*}

The MagE, UVES, and HIRES spectra cover several indicators of youth for mid-M dwarfs. For this paper, we have measured Li absorption, H$\alpha$ emission, and the Na~I index in both stars. The results are summarized in Table~\ref{tab:meas}. We note that both components show very strong Li absorption ($>$380~m\AA) and H$\alpha$ emission. 
{The presence of strong Li absorption is in agreement with other $\beta$~Pic moving group members of similar spectral types (see \citealt{Binks:2013}). The smaller equivalent width for LDS~5606A compared to B suggests 
that it may be close to the Li depletion boundary and that the age of the system cannot be significantly older than $\sim$20~Myr. 
}
{As further indication of a young age}, numerous emission lines are evident in the spectra (see Fig.~\ref{fig:spec}), including forbidden OI emission at 6300\AA\ (air). These are suggestive of ongoing accretion onto the central stars along with OH photodissociation in the disk and are discussed in more detail in Zuckerman et al.\ (submitted to ApJ). Although neither the MagE or UVES spectra could be telluric corrected, the Na~I index, a measure of surface gravity \citep{Lawson:2009}, has values of 1.16--1.18 and agrees well with those of other $\beta$~Pic members (see, for example, Fig.~5 in \citealt{Riedel:2014}). The numerous emission lines, strong Li absorption, and weak Na~I index all suggest a youthful age for this system.

At near-IR wavelengths, we measured the FeH$_z$, K$I_J$, and $H$-cont gravity indices from \citet{Allers:2013} and summarize our measurements in Table~\ref{tab:spex}. Unfortunately, the gravity scores in Table~9 of \citet{Allers:2013} only start at M6 and both components of LDS~5606 are M5. Nevertheless, extrapolating to M5 suggests that both components have very low surface gravity, consistent with the youth features observed in the optical spectra.

In addition to spectroscopic features of youth, both systems exhibit strong UV emission which led to their initial selection. LDS~5606B, in particular, had such strong UV emission that its NUV--W1 color fell outside the standard GALNYSS selection criteria and would not have been identified had LDS~5606A not been recovered. 
Both systems also show WISE infrared excesses similar to those identified around TWA~30AB, 31, 32, 33, and 34 (\citealt{Schneider:2012a,Schneider:2012b}; see also Fig.~\ref{fig:irexcess}). These excesses indicate the presence of warm circumstellar dust surrounding both stars (Zuckerman et al., submitted to ApJ).

Neither LDS~5606 star is detected in X-rays in the ROSAT All-Sky Survey (RASS), nor have Chandra or XMM observations been carried out. Assuming a RASS flux limit of $2\times10^{-13}$~ergs~cm$^{-2}$~s$^{-1}$ \citep{Schmitt:1995}, we estimate each component of LDS~5606 has log~$L_X/L_{bol}\lesssim-2.8$. This is just above the saturation limit observed for active M-dwarfs ($\sim$--3; see Fig.~5 in \citealt{Riaz:2006}). 
We note that TWA 30A was detected in the RASS (log~$L_X/L_{bol}=-3.34$; \citealt{Looper:2010a}) and expect that if LDS~5606 had been closer it would likely have been detected by ROSAT.

\subsection{Kinematics} \label{kinematics}

\begin{figure*}[htb]
\begin{center}
\includegraphics[width=17cm,angle=0]{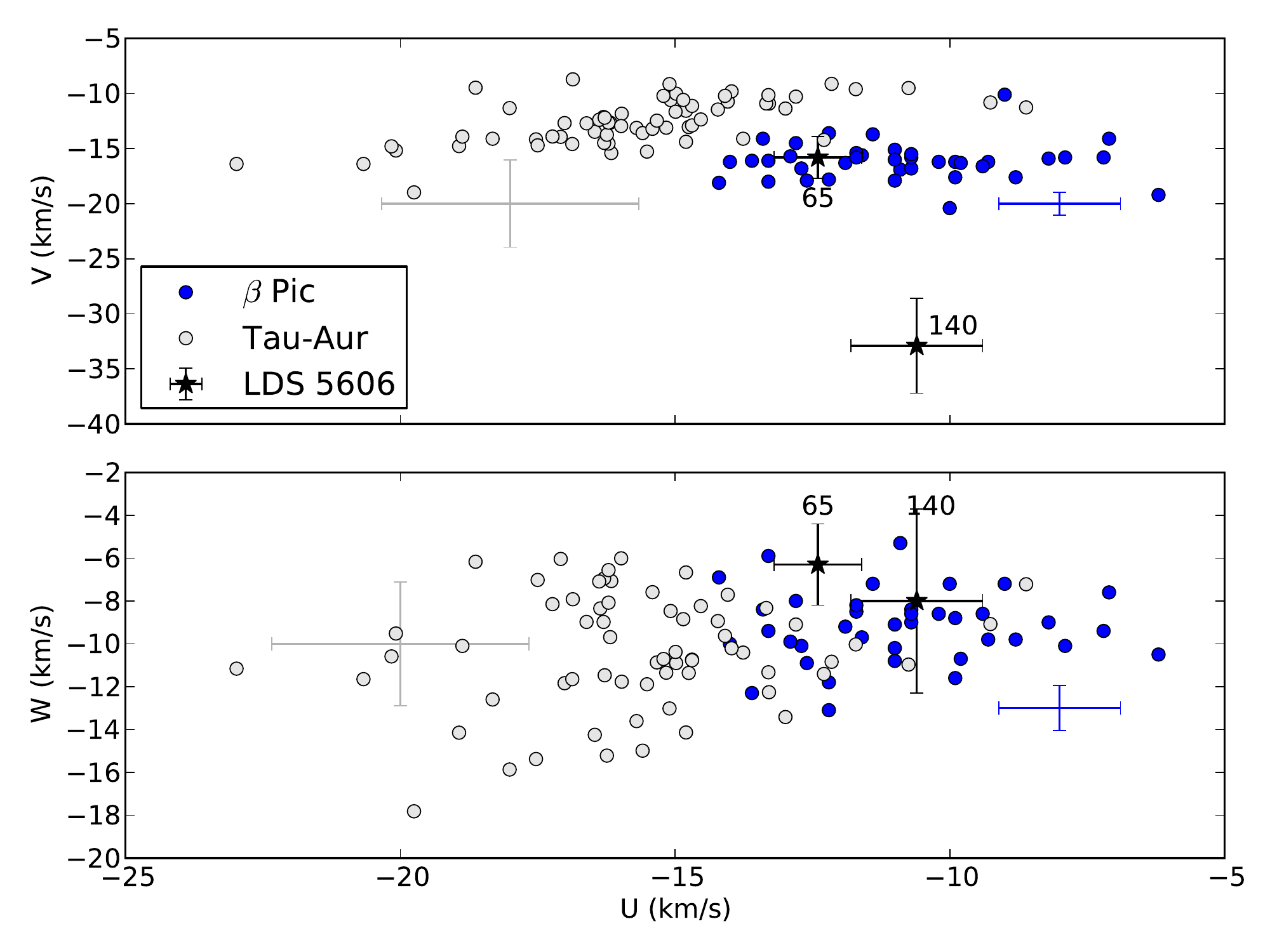}
\end{center}
\caption{Average UVW velocities of LDS~5606 at two distances (65, 140~pc) compared to $\beta$~Pic and Tau-Aur members \citep{Malo:2013,Bertout:2006}. Blue and grey error bars show the average uncertainty in the group member's UVW.}
\label{fig:uvw}
\end{figure*}

As previously mentioned and listed in Table~\ref{tab:summary}, we measure radial velocities and estimate a distance for LDS~5606.
The radial velocities are in good agreement with the convergent point and BANYAN estimates.
BANYAN can take as input the measured radial velocities to return a revised membership probability of 88\% for both components. With the new BANYAN~II tool \citep{Gagne:2014}, and selecting the option for young objects with priors set to unity, a $\beta$~Pic membership likelihood of 99\% is returned.
This is reflected by the system's UVW velocities, which we illustrate in Fig.~\ref{fig:uvw} for several distances for LDS~5606 and compare to known $\beta$~Pic and Tau--Aur members (see below).  

To test for possible membership in other groups, we computed UVW velocities with a range of distances from 50 to 150~pc in 1~pc steps. We find a good match to the $\beta$~Pic moving group UVW velocity at a distance of $\sim$66~pc, in good agreement with the BANYAN and convergent point results. However, the UVW velocities at a distance of $\sim$90~pc are also a good match to those of the Columba Association (\citealt{Torres:2008}). The convergent point \citep{Rodriguez:2013} returns a similar high likelihood of membership in the 30 Myr-old Columba Association of $\sim$70\%, with predicted radial velocity of 15.1~km/s and distance of 95~pc. On the other hand, BANYAN \citep{Malo:2013} returns a Columba membership likelihood of only 5\%, with similar radial velocity and distance requirements (14.9~km/s, 83~pc). 
Inclusion of the radial velocity only increases the BANYAN Columba likelihood to 11\%. 
If LDS~5606 were to be a Columba member, then at 90 pc, its $M_K$ would place it above various younger $\beta$ Pic members with similar V--K that are plotted in Fig.~\ref{cmd}.  Thus, LDS~5606 is unlikely to be a Columba member.

The distance of LDS~5606 (see Table~\ref{tab:summary}) places it somewhat removed from most known $\beta$~Pic members. 
However, while the $\beta$~Pic moving group is currently sparsely sampled at distances $>$50~pc, recent studies are identifying more candidates at similar distances as LDS~5606 (\citealt{Schlieder:2012c}, Vican et al.\ in prep.).
Nevertheless, it is worthwhile to consider if LDS~5606 may instead be a younger, more distant star. In particular, the Taurus-Auriga (Tau-Aur) star forming region at a distance of 140~pc \citep{Bertout:2006,Rebull:2011} has members in close proximity in the sky to LDS~5606. 
Tau-Aur members within 8 degrees of LDS~5606 have average proper motions of $3\pm9$, $-15\pm9$~mas/yr and average radial velocities of $17.6\pm1.3$~km/s \citep{HBC, Rebull:2011}. 
The proper motion and radial velocities of these Tau-Aur members differ from the measurements for LDS~5606.
Furthermore, the K-band magnitude of LDS~5606 is $\sim$1--2 magnitudes fainter than Tau-Aur members of similar colors suggesting that if it were at a distance of 140~pc, it would be underluminous compared to other stars of the same spectral type \citep{Rebull:2011}.
Finally, the UVW velocities of LDS~5606 at the average Tau-Aur distance of 140~pc are --11, --33, --8~km/s. These are markedly different from the UVW velocities listed in \citet{Bertout:2006}.
Figure~\ref{fig:uvw} compares the UVW of LDS~5606 at 65 and 140~pc with members of $\beta$~Pic \citep{Malo:2013} and Tau-Aur \citep{Bertout:2006}. There is no distance for which the UVW of LDS~5606 match those of Tau-Aur.
Given these considerations, membership of LDS~5606 in Tau-Aur is unlikely.

\subsection{Comparison with TWA~30}

LDS~5606 shows remarkable similarity to TWA~30AB \citep{Looper:2010a,Looper:2010b}.
Both systems have stars of similar spectral type ($\sim$M5) and are widely separated: $\sim$1700~AU for LDS~5606 and $\sim$3400~AU for TWA~30. TWA~30, as a member of the TW~Hya Association, is younger ($\sim$8~Myr) than LDS~5606. Both LDS~5606 and TWA~30 host warm dust disks around each component (Figure~\ref{fig:irexcess}).
The optical spectra of TWA~30AB, and LDS~5606A show numerous emission lines. However, with the exception of some [O$I$] emission, the emission lines seen in LDS~5606A are permitted transitions while forbidden transitions dominate the TWA~30 spectra (Zuckerman et al., submitted to ApJ).

\citet{Looper:2010a,Looper:2010b} interpret the TWA 30 spectra as due to a disk seen edge-on in the case of TWA 30B and nearly edge-on in the case of TWA 30A. The many forbidden lines in the optical spectra of both TWA~30 stars are produced in strong outflows nearly in the plane of the sky. The edge-on disks obscure the regions near the base of the accretion flows, regions that are responsible for the strong permitted lines expected in the magnetospheric model for disk accretion seen in many T Tauri stars.
The disks around LDS~5606, in contrast, are likely to be nearly face-on thus enabling the optical helium and hydrogen lines to be seen strongly in emission. 
Furthermore, neither TWA~30A or B are detected in GALEX whereas both components in LDS~5606 are. The edge-on disk may be blocking all UV emission from TWA~30, yet this emission is unobscured in LDS~5606, a scenario likely similar to that of the face-on disk of TW~Hya itself (see \citealt{Rodriguez:2011}). In contrast, TWA~30A is detected in X-rays with ROSAT, whereas TWA 30B and both components of LDS~5606 are not. \citet{Looper:2010b} attribute the TWA~30B X-ray non-detection to the edge-on disk. The non-detection for LDS~5606 is likely due to the greater distance of this pair.
The general absence of forbidden line emission indicates that any outflows at LDS~5606 are much weaker than at the TWA 30 stars (Zuckerman et al., submitted to ApJ).

\section{Conclusion}

LDS~5606 is a widely separated (26$''$) binary system consisting of two M5 dwarfs.
These stars were identified as part of the ongoing GALNYSS program \citep{Rodriguez:2013}, which seeks to identify low-mass members to nearby young moving groups by virtue of their strong ultraviolet emission.
Our kinematic and spectroscopic analysis of LDS~5606 place it as a member of the $\sim$20~Myr-old $\beta$~Pic moving group.
We estimate a kinematic distance of 65~pc, implying a projected separation of $\sim$1700~AU.
Features of youth are seen in the spectra of both stars. In particular, LDS~5606A shows numerous strong emission lines, especially of hydrogen and helium. 

Both components of LDS~5606 have IR excess at WISE wavelengths suggestive of warm circumstellar dusk disks.
As such, the properties of this binary are analogous to the younger ($\sim$8~Myr) TWA 30 binary \citep{Looper:2010a,Looper:2010b}. Both systems exhibit strong emission lines and host circumstellar disks, although the different orientation of these disks may account for the different emission lines observed and for the lack of GALEX detections for TWA~30.
To date, only a handful of binary stars located far from molecular clouds are known to host disks around each component.
Among stars with ages $\sim$8~Myr or older, LDS~5606 is one of only 4 binary or multiple systems known to host a pair of disks (TWA~30:\ \citealt{Looper:2010a,Looper:2010b}; HD223352:\ \citealt{Phillips:2011}; Fomalhaut:\ \citealt{Kennedy:2014}; and LDS~5606).
The unusual properties of the disks around the components of LDS~5606 and the accretion and photodissociation signatures in the spectra are discussed in detail in Zuckerman et al.\ (submitted to ApJ).
Identification of systems like LDS~5606 show that additional outstanding members of these young moving groups may yet be found.

\begin{acknowledgements}
We thank our referee, Eric Mamajek, for a detailed review that strengthened this paper.
This paper includes data gathered with the 6.5 meter Magellan Telescopes located at Las Campanas Observatory, Chile (CNTAC program CN2013A-135); the Infrared Telescope Facility, which is operated by the University of Hawaii under Cooperative Agreement no.\ NNX-08AE38A with the National Aeronautics and Space Administration, Science Mission Directorate, Planetary Astronomy Program; the Very Large Telescope under ESO program ID~092.C--0203(A); and the W.M. Keck Observatory, which is operated as a scientific partnership among the California Institute of Technology, the University of California and the National Aeronautics and Space Administration. 
The Keck Observatory was made possible by the generous financial support of the W.M. Keck Foundation.
The authors wish to recognize and acknowledge the very significant cultural role and reverence that the summit of Mauna Kea has always had within the indigenous Hawaiian community.  We are most fortunate to have the opportunity to conduct observations from this mountain.
D.R.R. acknowledges support from FONDECYT grant 3130520.
J.K.F. acknowledges support under NSF IRFP grant 0965192.
This research was supported in part by NASA ADAP grant NNX12AH37G to the Rochester Institute of Technology
and to UCLA, and by an NSF pre-doctoral fellowship to L.V.
\end{acknowledgements}


\begin{thebibliography}{}

\bibitem[Allers \& Liu(2013)]{Allers:2013} Allers, K.~N., \& Liu, M.~C.\ 2013, \apj, 772, 79
\bibitem[Baraffe et al.(1998)]{Baraffe:1998} Baraffe, I., Chabrier, G., Allard, F., \& Hauschildt, P.~H.\ 1998, \aap, 337, 403
\bibitem[Bertout \& Genova(2006)]{Bertout:2006} Bertout, C., \& Genova, F.\ 2006, \aap, 460, 499 
\bibitem[Binks \& Jeffries(2013)]{Binks:2013} Binks, A.~S., \& Jeffries, R.~D.\ 2013, \mnras, L192 
\bibitem[Bochanski et al.(2007)]{Bochanski:2007} Bochanski, J.~J., West, A.~A., Hawley, S.~L., \& Covey, K.~R.\ 2007, \aj, 133, 531
\bibitem[Bochanski et al.(2009)]{Bochanski:2009} Bochanski, J.~J., Hennawi, J.~F., Simcoe, R.~A., et al.\ 2009, \pasp, 121, 1409
\bibitem[Cutri et al.(2003)]{Cutri:2003} Cutri, R.~M., et al.\ 2003, The IRSA 2MASS All-Sky Point Source Catalog, NASA/IPAC Infrared Science Archive.~http://irsa.ipac.caltech.edu/applications/Gator/
\bibitem[Cushing et al.(2004)]{Cushing:2004} Cushing, M.~C., Vacca, W.~D., \& Rayner, J.~T.\ 2004, \pasp, 116, 362 
\bibitem[Cruz \& Reid(2002)]{Cruz:2002} Cruz, K.~L., \& Reid, I.~N.\ 2002, \aj, 123, 2828
\bibitem[de Bruijne(2012)]{deBruijne:2012} de Bruijne, J.~H.~J.\ 2012, \apss, 341, 31 
\bibitem[Dekker et al.(2000)]{Dekker:2000} Dekker, H., D'Odorico, S., Kaufer, A., Delabre, B., \& Kotzlowski, H.\ 2000, \procspie, 4008, 534 
\bibitem[Ducourant et al.(2014)]{Ducourant:2014} Ducourant, C. et al., 2014, A\&A, in press (arXiv:1401.1935)
\bibitem[Faherty et al.(2011)]{Faherty:2011} Faherty, J.~K., Burgasser, A.~J., Bochanski, J.~J., et al.\ 2011, \aj, 141, 71
\bibitem[Fischer \& Marcy(1992)]{Fischer:1992} Fischer, D.~A., \& Marcy, G.~W.\ 1992, \apj, 396, 178
\bibitem[Gagn{\'e} et al.(2013)]{Gagne:2014} Gagn{\'e}, J., Lafreni{\`e}re, D., Doyon, R., Malo, L., \& Artigau, {\'E}.\ 2013, arXiv:1312.5864 
\bibitem[Herbig \& Bell(1995)]{HBC} Herbig, G.~H., \& Bell, K.~R.\ 1995, VizieR Online Data Catalog, 5073, 0
\bibitem[Jordi et al.(2010)]{Jordi:2010} Jordi, C., Gebran, M., Carrasco, J.~M., et al.\ 2010, \aap, 523, A48
\bibitem[Kennedy et al.(2014)]{Kennedy:2014} Kennedy, G.~M., Wyatt, M.~C., Kalas, P., et al.\ 2014, \mnras, 438, L96
\bibitem[Kastner et al.(2011)]{Kastner:2011} Kastner, J.~H., Sacco, G.~G., Montez, R., et al.\ 2011, \apjl, 740, L17
\bibitem[Lawson et al.(2009)]{Lawson:2009} Lawson, W.~A., Lyo, A.-R., \& Bessell, M.~S.\ 2009, \mnras, 400, L29
\bibitem[Looper et al.(2010a)]{Looper:2010a} Looper, D.~L., Mohanty, S., Bochanski, J.~J., et al.\ 2010, \apj, 714, 45
\bibitem[Looper et al.(2010b)]{Looper:2010b} Looper, D.~L., Bochanski, J.~J., Burgasser, A.~J., et al.\ 2010b, \aj, 140, 1486 
\bibitem[Luhman et al.(2003)]{Luhman:2003} Luhman, K.~L., Stauffer, J.~R., Muench, A.~A., et al.\ 2003, \apj, 593, 1093
\bibitem[Luhman \& Muench(2008)]{Luhman:2008} Luhman, K.~L., \& Muench, A.~A.\ 2008, \apj, 684, 654
\bibitem[Luyten(1997)]{Luyten:1997} Luyten, W.~J.\ 1997, VizieR Online Data Catalog, 1130, 0
\bibitem[Malo et al.(2013)]{Malo:2013} Malo, L., Doyon, R., Lafreni{\`e}re, D., et al.\ 2013, \apj, 762, 88 
\bibitem[Mamajek et al.(2013)]{Mamajek:2013} Mamajek, E.~E., Bartlett, J.~L., Seifahrt, A., et al.\ 2013, \aj, 146, 154
\bibitem[Marshall et al.(2008)]{Marshall:2008} Marshall, J.~L., Burles, S., Thompson, I.~B., et al.\ 2008, \procspie, 7014, 169
\bibitem[Martin et al.(2005)]{Martin:2005} Martin, D.~C., et al.\ 2005, \apjl, 619, L1 
\bibitem[Mason et al.(2001)]{Mason:2001} Mason, B.~D., Wycoff, G.~L., Hartkopf, W.~I., Douglass, G.~G., \& Worley, C.~E.\ 2001, \aj, 122, 3466 
\bibitem[Monin et al.(2007)]{Monin:2007} Monin, J.-L., Clarke, C.~J., Prato, L., \& McCabe, C.\ 2007, Protostars and Planets V, 395
\bibitem[Murphy et al.(2012)]{Murphy:2012} Murphy, S.~J., Lawson, W.~A., \& Bessell, M.~S.\ 2012, \mnras, 424, 625
\bibitem[Nidever et al.(2002)]{Nidever:2002} Nidever, D.~L., Marcy, G.~W., Butler, R.~P., Fischer, D.~A., \& Vogt, S.~S.\ 2002, \apjs, 141, 503
\bibitem[Pecaut \& Mamajek(2013)]{Pecaut:2013} Pecaut, M.~J., \& Mamajek, E.~E.\ 2013, \apjs, 208, 9 
\bibitem[Phillips(2011)]{Phillips:2011} Phillips, N. M. 2011, PhD thesis, The University of Edinburgh
\bibitem[Rayner et al.(2003)]{Rayner:2003} Rayner, J.~T., Toomey, D.~W., Onaka, P.~M., et al.\ 2003, \pasp, 115, 362 
\bibitem[Rebull et al.(2011)]{Rebull:2011} Rebull, L.~M., Koenig, X.~P., Padgett, D.~L., et al.\ 2011, \apjs, 196, 4 
\bibitem[Reid et al.(1995)]{Reid:1995} Reid, I.~N., Hawley, S.~L., \& Gizis, J.~E.\ 1995, \aj, 110, 1838 
\bibitem[Riaz et al.(2006)]{Riaz:2006} Riaz, B., Gizis, J.~E., \& Harvin, J.\ 2006, \aj, 132, 866 
\bibitem[Riedel et al.(2014)]{Riedel:2014} Riedel, A.~R., Finch, C.~T., Henry, T.~J., et al.\ 2014, arXiv:1401.0722
\bibitem[Rodriguez et al.(2011)]{Rodriguez:2011} Rodriguez, D.~R., Bessell, M.~S., Zuckerman, B., \& Kastner, J.~H.\ 2011, \apj, 727, 62 
\bibitem[Rodriguez \& Zuckerman(2012)]{Rodriguez:2012} Rodriguez, D.~R., \& Zuckerman, B.\ 2012, \apj, 745, 147 
\bibitem[Rodriguez et al.(2013)]{Rodriguez:2013} Rodriguez, D.~R., Zuckerman, B., Kastner, J.~H., et al.\ 2013, \apj, 774, 101 
\bibitem[Roeser et al.(2010)]{Roeser:2010} Roeser, S., Demleitner, M., \& Schilbach, E.\ 2010, \aj, 139, 2440 
\bibitem[Schlieder et al.(2012)]{Schlieder:2012c} Schlieder, J.~E., L{\'e}pine, S., \& Simon, M.\ 2012, \aj, 144, 109 
\bibitem[Schmitt et al.(1995)]{Schmitt:1995} Schmitt, J.~H.~M.~M., Fleming, T.~A., \& Giampapa, M.~S.\ 1995, \apj, 450, 392 
\bibitem[Schneider et al.(2012a)]{Schneider:2012a} Schneider, A., Melis, C., \& Song, I.\ 2012a, \apj, 
754, 39 
\bibitem[Schneider et al.(2012b)]{Schneider:2012b} Schneider, A., Song, I., Melis, C., Zuckerman, B., \& 
Bessell, M.\ 2012b, \apj, 757, 163 
\bibitem[Shkolnik et al.(2011)]{Shkolnik:2011} Shkolnik, E.~L., Liu, M.~C., Reid, I.~N., Dupuy, T., \& Weinberger, A.~J.\ 2011, \apj, 727, 6 
\bibitem[Siess et al.(2000)]{Siess:2000} Siess, L., Dufour, E., \& Forestini, M.\ 2000, \aap, 358, 593 
\bibitem[Teixeira et al.(2008)]{Teixeira:2008} Teixeira, R., Ducourant, C., Chauvin, G., et al.\ 2008, \aap, 489, 825
\bibitem[Torres et al.(2008)]{Torres:2008} Torres, C.~A.~O., Quast, G.~R., Melo, C.~H.~F., \& Sterzik, M.~F.\ 2008, Handbook of Star Forming Regions, Volume II, 757 
\bibitem[Vacca et al.(2003)]{Vacca:2003} Vacca, W.~D., Cushing, M.~C., \& Rayner, J.~T.\ 2003, \pasp, 115, 389 
\bibitem[Vogt et al.(1994)]{Vogt:1994} Vogt, S.~S., Allen, S.~L., Bigelow, B.~C., et al.\ 1994, \procspie, 2198, 362 
\bibitem[Wright et al.(2010)]{Wright:2010} Wright, E.~L., Eisenhardt, P.~R.~M., Mainzer, A.~K., et al.\ 2010, \aj, 140, 1868 
\bibitem[Zacharias et al.(2004)]{Zacharias:2004} Zacharias, N., Monet, D.~G., Levine, S.~E., et al.\ 2004, Bulletin of the American Astronomical Society, 36, 1418 
\bibitem[Zacharias et al.(2012)]{Zacharias:2012} Zacharias, N., Finch, C.~T., Girard, T.~M., et al.\ 2012, VizieR Online Data Catalog, 1322, 0 
\bibitem[Zuckerman \& Song(2004)]{ZS04} Zuckerman, B., \& Song, I.\ 2004, \araa, 42, 685

\end{thebibliography}
\end{document}